\documentclass{article}
\usepackage{epsfig}
\usepackage{amssymb}
\newcommand{\bfr}{\begin{flushright}}
\newcommand{\efr}{\end{flushright}}
 
\begin{document}
\title{Quantum Corrections to Scattering Amplitude in Conical Space-time
%\thanks{Presented at ...}%
}
\author{Kiyoshi Shiraishi\\
%\address{
Akita Junior College, Shimokitade-Sakura, Akita-shi, \\Akita 010,
Japan%}
}
\date{Journal of the Korean Physical Society. {\bf 25} (1992) pp.
192--195 }
\maketitle
\begin{abstract}
It is known that the vacuum polarization of zero-point field arises
around a conical singularity generated by an infinite, straight
cosmic string. In this paper we study quantum electromagnetic
corrections to the gravitational Aharonov-Bohm effect around a cosmic
string. We find the scattering amplitude from a conical defect for
charged Klein-Gordon field.
\end{abstract}
%\PACS{}

%%%%%%%%%%%%%%%%%%%%%%%%%%%%%%%%%%%%%%%%%%%
\section{Introduction}
%%%%%%%%%%%%%%%%%%%%%%%%%%%%%%%%%%%%%%%%%%%
In the past few years, quantum theory on a cone has
attracted much attention. One of the motivations, besides theoretical
interest in $2+1$ dimensional field theory, is the application to
quantum field theory around cosmic strings. An idealized cosmic
string, which is infinitely thin, can be represented by a conical
singularity of space-time. Quantum effect around cosmic strings has
been studied by many authors including Dowker \cite{1}, Frolov
\cite{2}, Linet \cite{3} and Smith.\cite{4}

The scattering by a cosmic string or a conical singularity has also
been investigated recently.\cite{5} This non-trivial scattering is
often dubbed as the gravitational Aharonov-Bohm (AB) effect.

It is interesting to consider ``radiative'' corrections to
the gravitational AB scattering, which arise due to quantum
fluctuations of all interacting field around cosmic strings.

In the present paper, we consider vacuum fluctuations
of the electromagnetic field. Then, charged particles
scattered by a cosmic string undergo quatum corrections
even if the cosmic string has no charge and no classical
field strength.

The organization of this paper is as follows: In section
2, the vacuum polarization of gauge field is obtained.
In section 3, we derive the quantum-mechanically corrected
Klein-Gordon equation for a charged scalar field. The scattering
amplitude is calculated in section 4. Section 5 is devoted to
conclusion.

%%%%%%%%%%%%%%%%%%%%%%%%%%%%%%%%%%%%%%%%%%%
\section{The vacuum polarization of electromagnetic field}
%%%%%%%%%%%%%%%%%%%%%%%%%%%%%%%%%%%%%%%%%%%
In this section, we compute the vacuum value of
$\langle A_\mu A^\mu\rangle$ around a infinite straight cosmic
string, where $A_\mu$ is the electromagnetic field. Connection to an
effective wave equation for a charged scalar field is presented in the
succeeding section. We first anticipate that $\langle A_\mu
A^\mu\rangle\propto r^{-2}$, since there is no dimensional quantity
other than the distance from the string.

A comment on gauge invariance is in order. Classically
the presence of the cosmic string does not break the
gauge invariance. At the one-loop level, however, one
can see the breakdown of gauge Symmetry near the cosmic string, due
to the mass-like term $e^2\langle\phi^2\rangle A_\mu A^\mu$ in the
scalar QED; here the vacuum expectation value $\langle\phi^2\rangle$
depends on the distance from the cosmic string. Thus the non-zero
expectation value of $A_\mu A^\mu$ does not give rise to a real
conflict. Note also that we often encounter the expression $\langle
A_\mu A^\mu\rangle$ in the discussion on the Coleman-Weinberg
potential (or radiative gauge-symmetry breaking) in scalar QED at zero
and finite temperature.\cite{12}

We assume the following line element around the
idealized cosmic string which lies along the $z$ axis:
\begin{equation}
ds^2=-dt^2+dz^2+dr^2+(r^2/\nu^2)d\theta^2\,,
\label{2.1}
\end{equation}
where $\nu$ is related to the mass density of the string $\mu$
by $\nu^{-1}=1-4\pi G\mu$. Working in the Coulomb gauge, we
can denote the gauge field in the form of normal-mode
expansion, such as
\begin{equation}
A_i=\sum_{s=1,2} a^s A^s_i(r,\theta) e^{ik_s z-i\omega_s t}+h.c.   
\label{2.2}
\end{equation}
Mode functions $A^s_i$ are the solutions for the Maxwell
equations. In the coordinate system represented by eq.~(\ref{2.1}), we
find the following set of mode functions:
\begin{eqnarray}
A^1_z&=&\{i(l/\omega) J_{\nu n}(lr) \sin n\theta\,, 
i(l/\omega) J_{\nu n}(lr) \cos n\theta\}\,, 
\label{2.3-1} \\
A^1_r&=&\{i(k/2\omega) (J_{\nu n+1}(lr)-J_{\nu n-1}(lr)) \sin
n\theta\,, \nonumber \\
& &i(k/2\omega) (J_{\nu n+1}(lr)-J_{\nu n-1}(lr)) \cos
n\theta\}\,,
\label{2.3-2}\\
A^1_\theta&=&\{-(k/2\omega)(r/\nu)(J_{\nu n+1}(lr)+J_{\nu n-1}(lr)) \sin
n\theta\,,\nonumber \\
& &-(k/2\omega)(r/\nu)(J_{\nu n+1}(lr)+J_{\nu n-1}(lr)) \cos
n\theta\}\,, 
\label{2.3-3}\\ 
A^2_z&=&0\,,
\label{2.3-4} \\
A^2_r&=&\{(k/2)(J_{\nu n+1}(lr)+J_{\nu n-1}(lr)) \sin
n\theta\,,\nonumber  \\
& &(k/2)(J_{\nu n+1}(lr)+J_{\nu n-1}(lr)) \cos n\theta\}\,,
\label{2.3-5}\\
A^2_\theta&=&\{-(k/2)(r/\nu)J_{\nu n+1}(lr)-J_{\nu n-1}(lr)) \sin
n\theta\,,\nonumber \\
& & -(k/2)(r/\nu)(J_{\nu n+1}(lr)-J_{\nu n-1}(lr)) \cos n\theta\}\,,
\label{2.3-6}
\end{eqnarray}
where $J_m$ is the Bessel function, $n$ is positive integer and
$\omega^2=k^2+l^2$.

Using the mode functions, we can perform the calculation of
$\langle A_i A^i\rangle$ by mode-summation. After regularization, we
find that a finite portion of $\langle A_i A^i\rangle$ is given by
\begin{equation}
\langle A_i A^i\rangle=\frac{\nu^2-1}{24\pi^2r^2}\,.
\label{2.4}
\end{equation}

The quantity has been regularized so as to become
zero if $\nu=1$. (Note that this is twice of the real-scalar
contribution.\cite{4}) The mode expansion is also useful to
investigale properties of charged fields around a cosmic
string.\cite{6}

In the next section, we consider the modification of
field equation due to the vacuum field $\langle A_i A^i\rangle$.

%%%%%%%%%%%%%%%%%%%%%%%%%%%%%%%%%%%%%%%%%%%
\section{Modified Klein-Gordon equation for a charged scalar}
%%%%%%%%%%%%%%%%%%%%%%%%%%%%%%%%%%%%%%%%%%%
Suppose that a minimally-coupled scalar field $\phi$ is governed by
Klein-Gordon equation, where the space-time derivative is replaced by
the covariant derivative including the gauge field.

In this paper, we assume that the cosmic string has
no classical electromagnetic field. Although general cosmic strings in
GUTs have fluxes in their core, their fluxes are not magnetic fluxes
we know, but are associated with gauge fields of broken symmetry. For
a superconducting cosmic string \cite{7} the analysis ought to be done
by some other ways, since the metric is more or less
deformed by electric currents. Therefore, we treat the
case with normal cosmic strings throughout this paper.

However, we take into account the vacuum expcctation value of the
electromagnetic field. Only the second moment $\langle A_i
A^i\rangle$ appears and the field equation is expressed as
\begin{equation}
\Box\phi-e^2\langle A_i A^i\rangle\phi-m^2\phi=0\,,
\label{3.1}
\end{equation}
where $e$ is the coupling constant and $m$ is the mass. The
term $e^2\langle A_i A^i\rangle$ behaves as a potential term.

%We set $m=0$ in the following analysis in this paper.

If we use the metric (\ref{2.1}) and substitute (\ref{2.2}) into
eq.~(\ref{3.1}), we obtain
\begin{equation}
\frac{1}{r}\frac{\partial}{\partial
r}r\frac{\partial\phi}{\partial
r}+\frac{\nu^2}{r^2}\frac{\partial^2\phi}{\partial\theta^2}-
\frac{\gamma^2}{r^2}\phi-m^2\phi=0\,,
\label{3.2}
\end{equation}
where $\gamma^2=e^2(\nu^2-1)/24\pi^2$.

A solution for the modified equation (\ref{3.2}) is written in the form
\begin{equation}
\phi(t,z,r,\theta)=J_{\alpha_n}(lr) e^{ikz-i\omega t+in\theta}\,,
\label{3.3}
\end{equation}
with $\alpha_n=(\nu^2n^2+\gamma^2)^{1/2}$ and $\omega^2=l^2+k^2+m^2$.

In the following section, we study the scattering problem by use of
this solution.

We set $m=0$ in the following analysis in this paper.

%%%%%%%%%%%%%%%%%%%%%%%%%%%%%%%%%%%%%%%%%%%
\section{Scattering amplitude in the presence of the quantum potential}
%%%%%%%%%%%%%%%%%%%%%%%%%%%%%%%%%%%%%%%%%%%
To treat the scattering by an infinite string, we consider
two-dimensional scattering, taking the $z$-component of the momentum
($k$ in (\ref{3.3})) to be zero. We have set
$m=0$.

We first review the scattering in two dimensions.\cite{5,8}
The wave function behaves asymptotically,
\begin{equation}
\phi\stackrel{r\rightarrow \infty}{\longrightarrow}e^{ilr\cos\theta}
+e^{i\pi/4}\frac{f(\theta)}{r^{1/2}}e^{ilr}\,,
\label{4.1}
\end{equation}
when a target sits at the origin. Here $f(\theta)$ is a scattering
amplitude, and a differential cross section (for a cosmic string, per
unit length) is given by $|f(\theta)|^2$. The phase in front of
$f(\theta)$ is chosen to simplify the expression of the optical theorem
in two dimensions \cite{5,8} (see later). In order to find the
scattering amplitude, we compare the asymptotic form of the solution
of the wave equation with (\ref{4.1}). We determine $f$ by matching the
coefficient of the in-going ``spherical'' wave ($\propto e^{-ilr}$).

The asymptotic form of the plane wave going in the
$x$-direction is given by
\begin{equation}
e^{ilr\cos\theta}\stackrel{r\rightarrow\infty}{\longrightarrow}
(2\pi lr)^{-1/2}e^{i\pi/4}\sum_{n=-\infty}^\infty
\{-i e^{ilr}+(-1)^n e^{-ilr}\} e^{in\theta}\,.
\label{4.2}
\end{equation}
Further if we assume that we can take the following asymptotic
behavior of the solution for the wave equation:
\begin{equation}
\phi\stackrel{r\rightarrow\infty}{\longrightarrow}
(2\pi lr)^{-1/2}e^{i\pi/4}\sum_{n=-\infty}^\infty
\{-i e^{ilr+2i\delta_n}+(-1)^n e^{-ilr}\} e^{in\theta}\,,
\label{4.3}
\end{equation}
then we find
\begin{equation}
f(\theta)=-i(2\pi l)^{-1/2}\sum_{n=-\infty}^\infty
(e^{2i\delta_n}-1) e^{in\theta}\,, 
\label{4.4}
\end{equation}
where $\delta_n$ is called as a phase shift.

Now, we will turn back to the present problem. The
wave equation to be considered is eq.~(\ref{3.2}) with $m=0$.
We regard $\gamma$ as an independent parameter of $\nu$ for a
later use. The mode solution (\ref{3.3}) behaves asymptotically as
\begin{equation}
\phi_n\stackrel{r\rightarrow\infty}{\longrightarrow}
(2\pi lr)^{-1/2}e^{i\pi/4}A_n
\{-i e^{ilr-i\pi(\alpha_n-n)}+(-1)^n e^{-ilr}\} e^{in\theta}\,,
\label{4.5}
\end{equation}
where $A_n=(-1)^n e^{-in\alpha_n/2}$.
Thus one can find
\begin{equation}
\delta_n=-(\pi/2)(\alpha_n-n)=-(\pi/2)\{(\nu^2n^2+\gamma^2)^{1/2}-n\}\,.
\label{4.6}
\end{equation}

If $\gamma^2$ is independent of $\nu$, $\delta_n$ is negative
when $\nu=1$ and $\gamma^2>0$.
Therefore we can say that the ``potential'' $\gamma^2/r^2$
gives rise to a repulsive force. As studied and stated
in ref.~\cite{5}, the scattering amplitude for $\nu\ne 1$ contains
delta functions. Namely, for $\gamma^2=0$, one can find \cite{5}
\begin{eqnarray}
& &(2\pi
l)^{1/2}f(\theta)=
\frac{\sin\{(\nu-1)\pi\}}{\cos\{(\nu-1)\pi\}-\cos\theta}
-i\pi
\sum_n\{\delta(\theta+(\nu-1)\pi-2\pi n)\nonumber \\
& &+\delta(\theta-(\nu-1)\pi-2\pi
n)-2\delta(\theta-2\pi n)\}\,.
\label{4.7}
\end{eqnarray}
This peculiarity originates from the conical space generates
``long-range force'' in some naive sense. We discuss the amplitude in
which such divergences are removed.\cite{5}

We will calculate
\begin{eqnarray}
(2\pi l)^{1/2}{\rm
Re}f(\theta)&=&\sum_n\sin(n\theta-\pi(\alpha_n-n))\,,
\label{4.8-1}\\
(2\pi l)^{1/2}{\rm
Im}f(\theta)&=&-\sum_n\cos(n\theta-\pi(\alpha_n-n))\,,
\label{4.8-2}
\end{eqnarray}
where $\theta+(\nu-1)\pi\ne 2\pi n$ and $\theta-(\nu-1)\pi\ne 2\pi n$
($n$: integer). We first compute the scattering amplitude for
$\nu=1$ and $\gamma\ne 0$ (we take $\gamma$ as an independent
parameter.). This is used for a check of general results. The
result obtained by numerical calculation is shown in Fig.~1. We examine
some approximation schemes.

%**************************************************************
\begin{figure}[ht]
\begin{center}
\includegraphics[width=6cm]{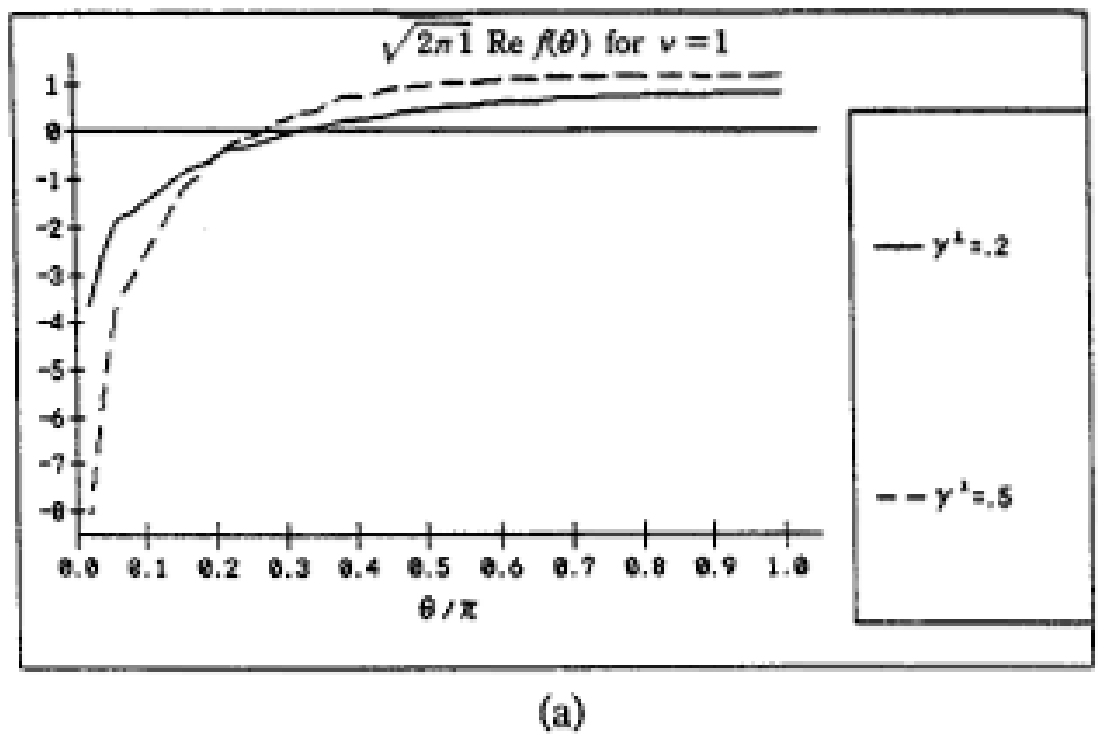}
\includegraphics[width=6cm]{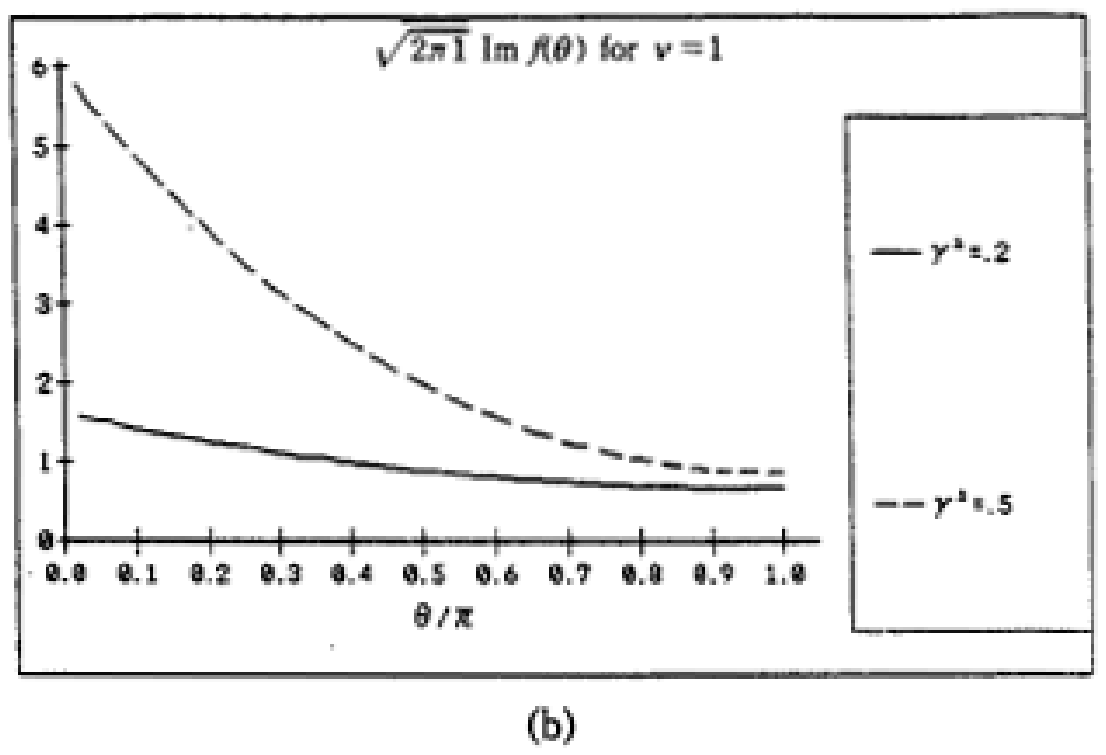}
\caption{The scattering amplitude for $\nu=1$ and $\gamma\ne 0$.
(a) $(2\pi l)^{1/2}{\rm Re} f(\theta)$, (b) $(2\pi l)^{1/2}{\rm Im}
f(\theta)$. The solid line stands for $\gamma^2=0.2$ and the broken line
for $\gamma^2=0.5$.}
\label{f1}
\end{center}
\end{figure}
%**************************************************************

The $S$-wave ($n=0$) contribution in the formula (\ref{4.4})
gives
\begin{eqnarray}
(2\pi l)^{1/2}{\rm
Re}f_0&=&-2\sin(\pi\gamma/2)\cos(\pi\gamma/2)\,,
\label{4.9-1}\\
(2\pi l)^{1/2}{\rm
Im}f_0&=&2\sin^2(\pi\gamma/2)\,,
\label{4.9-2}
\end{eqnarray}
and these seem to be a good estimation as the lowest order.

The optical theorem
\begin{equation}
{\rm Im} f(0)=(l/8\pi)^{1/2}\int_0^{2\pi}|f(\theta)|^2d\theta\,,
\label{4.10}
\end{equation}
is applicable for $\nu=1$ and one can confirm this at the
order in using the $S$-wave result (\ref{4.9-1}, \ref{4.9-2}). Note
that the amplitude does not contain divergences for $\nu=1$ (for
instance, delta functions cancel one another in (\ref{4.7})
when $\nu=1$.).

Although the potential has a short-range nature, the
Born approximation is not applicable because the singularity near
the origin is too strong.

Next we consider the case for a cosmic string. Here
we recover the relation $\gamma^2=e^2(\nu2-1)/24\pi^2$. We define
$\Delta f$ by the difference in the scattering amplitudes for
$\gamma^2=e^2(\nu2-1)/24\pi^2$ and $\gamma^2=0$; i.e., the contribution
of (\ref{4.7}) is subtracted. The numerical results are given in
Fig.~2. The value of ${\rm Im} \Delta f$ in the vicinity of $(\nu-1)\pi$
is not indicated because of the poor convergence in the
sum of the oscillating series. The value seems almost
constant.

%**************************************************************
\begin{figure}[ht]
\begin{center}
\includegraphics[width=6cm]{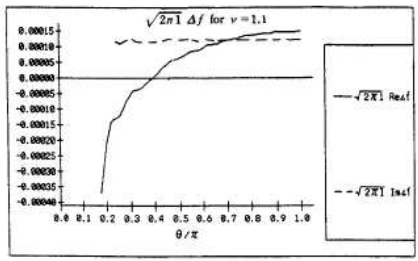}
\caption{The difference in the scattering amplitudes for $\gamma^2=e^2
(\nu^2-I)/24\pi^2$ and $\gamma^2=0$ in the case of $\nu-1=0.1$. We
set $e^2/4\pi=1/137$. The solid line stands for $(2\pi l)^{1/2}{\rm Re}
\Delta f$ and the broken line for  $(2\pi l)^{1/2}{\rm Im}
\Delta f$.}
\label{f2}
\end{center}
\end{figure}
%**************************************************************

The real part of $\Delta f$ exhibits a logarithmical divergence at
$\theta=(\nu-1)\pi$. The behavior of the amplitude is approximately
given by:
\begin{eqnarray}
& &(2\pi l)^{1/2}{\rm Re} \Delta
f=-2\sin(\pi\gamma/2)\cos(\pi\gamma/2)
\nonumber \\
& &\qquad+\frac{\pi\gamma^2}{2\nu}\ln
\left|4 \sin\frac{\theta+(\nu-1)\pi}{2}
\sin\frac{\theta-(\nu-1)\pi}{2}\right| +O(\gamma^4)\,,
\label{4.11-1}
\end{eqnarray}
\begin{eqnarray}
& &(2\pi l)^{1/2}{\rm Im} \Delta
f=2\sin^2(\pi\gamma/2)
\nonumber \\
& &\qquad+\frac{\pi\gamma^2}{2\nu}
\left\{\nu-1-
\left\{
\begin{array}{cc}
1 & (\theta<(\nu-1)\pi)\\
0 & ((\nu-1)\pi<\theta)
\end{array}
\right\}
\right\}
+O(\gamma^4)\,.
\label{4.11-2}
\end{eqnarray}
For a realistic cosmic string, $\nu-1$ may be too small for
us to detect the deviation in the scattering amplitude
unless the strength of the coupling become sufficiently
large.

%%%%%%%%%%%%%%%%%%%%%%%%%%%%%%%%%%%%%%%%%%%
\section{Conclusion}
%%%%%%%%%%%%%%%%%%%%%%%%%%%%%%%%%%%%%%%%%%%
In this paper we have derived the lowest-order vacuum quantum space
electromagnetic corrections to the scattering amplitude in a
conical background space-time. Inclusion of the quantum effect of
non-Abelian gauge fields around a cosmic string is an important
extension of the present work: it may be connected with non-Abelian
AB effect \cite{9} and the baryon decay/genesis mediated by cosmic
strings.\cite{10} It is also important to treat the effect of
self-interaction in a conical space for the Yang-Mills case.
(Self-interacting scalar fields around a cosmic string have been
studied in ref.~\cite{11}.).

%%%%%%%%%%%%%%%%%%%%%%%%%%%%%%%%%%%%%%%%%%%%%% 

%%%%%%%%%%%%%%%%%%%%%%%%%%%%%%%%%%%%%%%%%%%%%
\end{document}